\documentclass[nonacm]{acmart}

\usepackage{xcolor}
\usepackage{lmodern}
\usepackage{graphicx}

\usepackage{listings} 
\newcommand{\ic}[1]{{\tt #1}}

\begin{document}
\title{Discrete Event Simulation: It's Easy with SimPy!}
\author{Dmitry Zinoviev}
\address{Suffolk University, Boston}
\email{dzinoviev@suffolk.edu}
\date{February 2018}

\begin{abstract}
This paper introduces the practicalities and benefits of using SimPy,
a discrete event simulation (DES) module written in Python, for
modeling and simulating complex systems. Through a step-by-step
exploration of the classical Dining Philosophers Problem, we
demonstrate how SimPy enables the efficient construction of discrete
event models, emphasizing system states, transitions, and event
handling. We extend the scenario to introduce resources, such as
chopsticks, to model contention and deadlock conditions, and showcase
SimPy's capabilities in managing these scenarios. Furthermore, we
explore the integration of SimPy with other Python libraries for
statistical analysis, showcasing how simulation results inform system
design and optimization. The versatility of SimPy is further
highlighted through additional modeling scenarios, including resource
constraints and customer service interactions, providing insights into
the process of building, debugging, simulating, and optimizing models
for a wide range of applications. This paper aims to make DES
accessible to practitioners and researchers alike, emphasizing the
ease with which complex simulations can be constructed, analyzed, and
visualized using SimPy and the broader Python ecosystem.
\end{abstract}
\maketitle

\section{Introduction}
Computer modeling and simulation (M\&S) is the priceless art of
bringing to ``life'' systems and behaviors that would otherwise be
prohibitively expensive, unethical, or impossible to build, such as a
Mars rover, a centaur or a cruel autocratic regime based on slave
ownership.

An M\&S project consists of several standard
steps~\cite{Banks2009DiscreteEvent}:

\begin{enumerate}
\item First, an M\&S specialist builds a {\em model}. The most common
  model types are:
\begin{itemize}
\item Mathematical models that represent the relationships between the
  system variables as one or more linear or differential equations
  (Ohm's law is an example of a linear model), and
\item Discrete event models that represent the system as a collection
  of finite states and transitions between them caused by internal or
  external events. The simulation time ``jumps'' from one event to the
  next, and there is a timeless void between events.
\end{itemize}

\item Then, the model is {\em simulated} to obtain execution
  traces. The traces describe the evolution of some or all system
  variables over time.
\item The traces are often {\em visualized} because a picture is worth
  1,000 words.
\item {\em Optimization} is the next step if the project aims to
  discover the optimal values of the system parameters. The system
  model is repeatedly simulated during optimization with different
  parameters to minimize or maximize a predefined goal function. The
  optimization routine may use constrained or unconstrained functional
  optimization, curve fitting, simulated annealing, or brutal force.
\end{enumerate}

Once implemented, debugged, simulated, and optimized, a model can be
used to evaluate, build, and even replace the original system.

There is a plethora of general-purpose M\&S software, both commercial
and open source, such as \ic{OpenModelica}~\cite{openmodelica},
\ic{Ptolemy}~\cite{Ptolemaeus2014}, \ic{Simulink}~\cite{MATLAB}, and
even \ic{Simula}~\cite{kirkerud1989}---a complete computer programming
language designed to support simulation. This narrowly focused article
shows how to do simple discrete event simulation with
\ic{SimPy}~\cite{simpy}---an M\&S module written in pure Python.

\section{Installing \ic{SimPy}}

The easiest (and the only recommended) way to install
\ic{SimPy} is via \ic{pip}:

\begin{lstlisting}[language=bash]
pip install simpy
#> ...
#> Successfully installed simpy-4.1.1
\end{lstlisting}

At the time of writing, \ic{simpy-4.1.1} is the most recent version of
\ic{SimPy} and will be used for all examples.

To check if \ic{SimPy} was successfully installed, open a
Python shell and \ic{import simpy}.

Assuming you see no error messages, the next goal is finding a
suitable model and simulation system. Why not revisit the good old
Dining Philosophers?

\section{Dining Philosophers}

In brief, five outstanding Philosophers sit around a table with an
enormous dish of freshly made spaghetti at the center. Most of the
time, they think, but now and then, a Philosopher becomes hungry. At
that point, he collects two forks---left and right---and
starts eating the pasta. Having refueled himself, he puts down the
forks and returns to thinking.

Like all fine philosophers, the five dining philosophers barely make
ends meet and can afford only five forks, which prevents more than two
of them from eating simultaneously. In addition to being poor, the
philosophers are stubborn and uncooperative: once they pick up
one of the forks, they only put it down once they have a chance to
eat. Think what happens if all five guys pick up their left forks at
about the same time. (A hint: they starve.)

\begin{figure}[tb]\centering
  \includegraphics[width=.65\columnwidth]{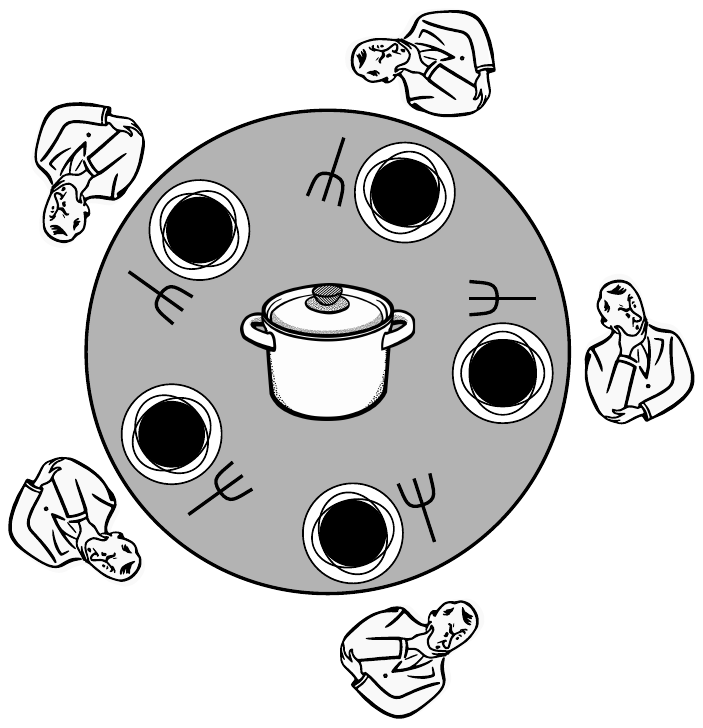}
  \caption{\label{table}The five dining philosophers}
  \Description{The five dining philosophers around the table.}
\end{figure}

Edsger Dijkstra introduced the dining philosophers problem in 1965,
and it has been a gold standard test framework for concurrent
algorithm designers~\cite{Silberschatz2018}. The problem is simple and
well-understood, which makes it an excellent candidate for an
introductory simulation project. The proposed solution models each
philosopher and the interactions between the philosophers and the
chopsticks.

From the discrete event simulation point of view, a philosopher is a
finite state machine with four states: \ic{Thinking} (the initial
state), \ic{Hungry}, \ic{Hungry-With-One-Chopstick}, and
\ic{Eating}. The diagram has no final state, as the philosophers are
never supposed to break the cycle.

\begin{figure}[tb]\centering
  \includegraphics[width=.5\columnwidth]{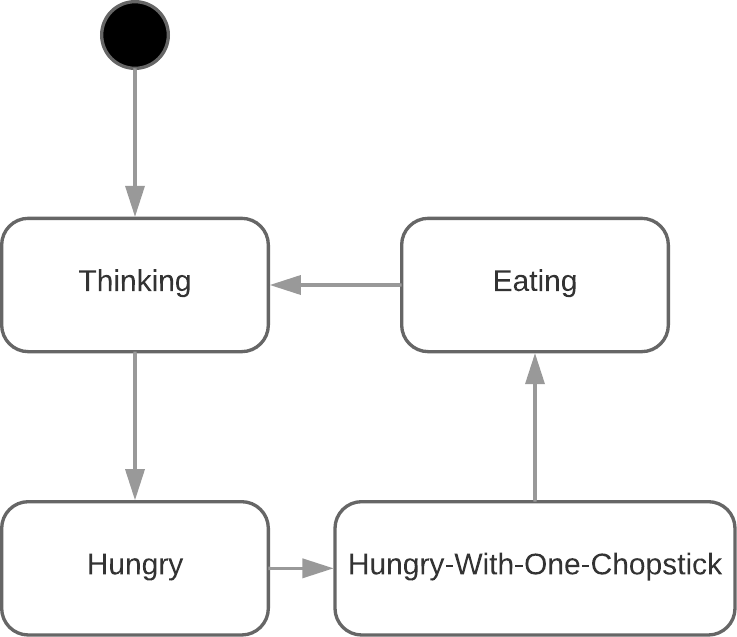}
  \caption{\label{philosopher-state}State transition diagram of a
    dining philosopher}
  \Description{State transition diagram of a
    dining philosopher.}
\end{figure}

The transitions between the states are caused by external or internal
events. A thinking philosopher becomes hungry {\em after some
  time}---on timeout. The time is based on some internal philosopher's
considerations that are unknown to us. It can be modeled as an
exponentially distributed random variable\footnote{It is a common and
dangerous mistake to treat times as normally distributed random
variables---in the first place, due to the unbounded nature of a
normal distribution.} with the mean value \ic{T0}. The transition from
the \ic{Eating} state to the \ic{Thinking} state is triggered
internally, too. It can be modeled as another exponentially
distributed random variable with the mean value \ic{T1}.

The other two transitions in the state transition diagram result from
external events: the availability of the chopsticks. A philosopher
becomes \ic{Hungry-With-One-Chopstick} when he manages to collect the
first chopstick, and he becomes \ic{Eating} when the second chopstick
becomes available.

Last, it is assumeed a philosopher spends some small but finite time
\ic{DT} in the \ic{Hungry-With-One-Chopstick} state even if the other
chopstick is readily available. If he picks up both chopsticks at
once, the system never ends up in a deadlock, which is practically
desirable---but not very exciting from the S\&M point of view.

\section{Simulating the ``Classical'' Dining Philosophers}

Enter \ic{SimPy}.

At the core of any \ic{SimPy} model, there is an environment
(\ic{simpy.Environment}). The environment, among other things,
provides functions for the continuous or step-by-step simulation of
the models, dispatches events, and track of the current simulation
time. A \ic{SimPy} program starts by importing the necessary modules
and creating a new simulation environment:

\begin{lstlisting}
import simpy
import random

env = simpy.Environment()
\end{lstlisting}

The environment will be passed to all other entities involved in the
model---for example, to the chopsticks.

Using the S\&M terminology, a chopstick is a {\em renewable
  resource}. It is managed by the environment and must be created
before the first use. During the simulation, a resource can be
requested, allocated, and released. \ic{SimPy} provides three types of
resources: ``vanilla'' resources (\ic{simpy.Resource},
\ic{simpy.PriorityResource}, \ic{simpy.PreemptiveResource}),
containers (\ic{simpy.Container}), and stores (\ic{simpy.Store},
\ic{simpy.FilterStore}, \ic{simpy.PriorityStore}). The first model of
a chopstick is the basic \ic{simpy.Resource}.

Each resource has a capacity that defines how many resource users can
share it. If the number of requests exceeds the capacity, \ic{SimPy}
blocks the outstanding requests until some users release the
resource. For sanitary and convenience reasons, a chopstick should not
be shared; its capacity is set to 1 (making it a {\em mutually
  exclusive} resource). Five chopsticks are created:

\begin{lstlisting}
N = 5
chopsticks = [simpy.Resource(env, capacity=1) for i in range(N)]
\end{lstlisting}

The next step is to define the philosophers. Unlike the chopsticks, the
philosophers are active. In \ic{SimPy}, active entities are known as
processes. (Not to be confused with the operating system processes!) A
process is a Python generator that yields discrete events. (If your
knowledge of the \ic{yield} keyword is rusty, refresh it before
reading further.) The simulation environment has a method
\ic{env.process()} that registers the processes. The following
code declares the process function \ic{run\_the\_party()} as a
Python class method and creates and registers an event generator after
initializing the class and instance variables. The purpose of
highlighting is explained later.

\begin{lstlisting}
class Philosopher():
    T0 = 10 # Mean thinking time
    T1 = 10 # Mean eating time
    DT = 1 # Time to pick the other chopstick

    def __init__(self, env, chopsticks, my_id, DIAG = False):
        self.env = env
        self.chopsticks = chopsticks
        self.id = my_id
	# START_HIGHLIGHT
        self.waiting = 0
        # END_HIGHLIGHT
	self.DIAG = DIAG
	# Register the process with the environment
        env.process(self.run_the_party())

    def get_hungry(self): # Request the resources
        yield # Do nothing so far

    def run_the_party(self): # Do everything...
        yield # ...but do nothing so far

    def diag(self, message): # Diagnostic routine
        if self.DIAG:
            print("P{} {} @{}".format(self.id, message, self.env.now))
\end{lstlisting}

The class constructor takes several parameters: the simulation environment, the chopstick resources, and the philosopher's id (for diagnostics). Also note that the current simulation time is available as an attribute of the environment, \ic{env.now}.

The philosophers are dysfunctional and do not yield any events, but they
can already be instantiated and assigned two chopsticks to each:

\begin{lstlisting}
philosophers = [Philosopher(env, (chopsticks[i], chopsticks[(i + 1) % N]), i) 
                for i in range(N)]
\end{lstlisting}

The event generator \ic{run\_the\_party()} yields events of three
types: timeout, request, and release. A call to \ic{env.timeout(tau)}
blocks the generator for \ic{tau} time units. Delays are used to model
the thinking and eating activities. A call to \ic{resource.request()}
blocks the generator until the \ic{resource} becomes available, marks
the resource as allocated to the caller, and returns the request
ID. Finally, a call to \ic{resource.release(rq)} deallocates the
\ic{resource} associated with the request id \ic{rq}. It does not
block the generator.

The following code is a translation of Figure~\ref{philosopher-state}
to the language of \ic{SimPy}. It consists of two methods that
implement the main process and a subprocess.

\begin{lstlisting}
class Philosopher():
    ...

    def get_hungry(self):
        # START_HIGHLIGHT
        start_waiting = self.env.now
        # END_HIGHLIGHT
        self.diag("requested chopstick")
        rq1 = self.chopsticks[0].request()
        yield rq1 
        
        self.diag("obtained chopstick")
        yield self.env.timeout(self.DT)

        self.diag("requested another chopstick")
        rq2 = self.chopsticks[1].request()
        yield rq2

        self.diag("obtained another chopstick")
        # START_HIGHLIGHT
        self.waiting += self.env.now - start_waiting
        # END_HIGHLIGHT
        return rq1, rq2

    def run_the_party(self):
        while True:
	    # Thinking
	    thinking_delay = random.expovariate(1 / self.T0)
            yield self.env.timeout(thinking_delay)

	    # Getting hungry
	    get_hungry_p = self.env.process(self.get_hungry())
            rq1, rq2 = yield get_hungry_p
	    
	    # Eating
	    eating_delay = random.expovariate(1 / self.T1)
            yield self.env.timeout(eating_delay)

	    # Done eating, put down the chopsticks
            self.chopsticks[0].release(rq1)
            self.chopsticks[1].release(rq2)
            self.diag("released the chopsticks")
\end{lstlisting}

The method \ic{get\_hungry()} is a subprocess---a process within a
process. It hides the technicalities of getting hungry (get one
chopstick, get another chopstick, etc.), letting me concentrate on the
party's conceptual flow. The main process is suspended until the
subprocess yields all events. Note that the subprocess method returns
the chopstick request handlers used later to release the resources.

The model is ready to run. It can be simulated step by step (or event by
event) by calling \ic{env.step()}, or for the first \ic{t} time units
by calling \ic{env.run(until=t)}, or until the simulation naturally
stops by calling \ic{env.run()} without any parameters. Here is a
sample output of the script:

\begin{lstlisting}
env.run()
#> P1 requested chopstick @0.10843721582414197
#> P1 obtained chopstick @0.10843721582414197
#> P1 requested another chopstick @1.108437215824142
#> P1 obtained another chopstick @1.108437215824142
#> P0 requested chopstick @6.642119333001387
#> P0 obtained chopstick @6.642119333001387
#> P0 requested another chopstick @7.642119333001387
#> P2 requested chopstick @7.898070765597094
#> ...
#> P2 obtained chopstick @45632.71748485687
#> P1 requested another chopstick @45632.721825172666
#> P4 requested another chopstick @45632.94176319329
#> P3 requested another chopstick @45633.53086921637
#> P2 requested another chopstick @45633.71748485687
\end{lstlisting}

The simulation suddenly stops at this point: the system is in a
deadlock state where each philosopher holds one chopstick and waits
for the other. There is no standard way to detect deadlocks in
\ic{SimPy}. An indirect solution is to look at the counts of requests
or actual requests granted for each resource (stored in the attributes
\ic{resource.count} and \ic{resource.users}, respectively). If at
least two counts are greater than zero and the simulation stops, the
system has likely been deadlocked:

\begin{lstlisting}
[f.count for f in chopsticks]
#> [1, 1, 1, 1, 1]
\end{lstlisting}

Incidentally, requests not yet granted can be found in the attribute
\ic{resource.queue}.

One of the well-known solutions to a deadlock is to allocate resources
in some globally defined order (say, in the order of their numerical
IDs or memory references). The philosophers will never starve if a
Python ID-based sorting is added to the second line of the class
initializer:

\begin{lstlisting}
self.chopsticks = sorted(chopsticks, key=id)
\end{lstlisting}

\section{Gathering and Visualizing Statistics}

The goal of an M\&S project is usually to verify the correctness of
the modeled system and get a sense of its performance. The most common
performance characteristics include waiting time, turnaround time,
response time, and resource utilization. As an exercise, let's
estimate waiting time---the time a process waits for the requested
resources (ready to run but not running).

Unlike other M\&S software, \ic{SimPy} does not provide tools for
gathering and visualizing simulation statistics---and, honestly, it
does not have to. As a part of the Python ecosystem, \ic{SimPy} can be
easily integrated with \ic{statistics}, \ic{NumPy}, \ic{Pandas},
\ic{SciPy}, \ic{matplotlib}, and other plotting, statistical, and
general number-crunching packages.

The mean waiting time of the philosophers is measured by instrumenting
the \ic{Philosopher} class with a time-counting variable
\ic{waiting}. The relevant instrumentation code is highlighted in the
listings above.

To compare the waiting times or any other performance metrics for
different systems (say, the dining parties with various numbers of
philosophers), a function is needed that prepares a model, makes a
simulation run, and returns the appropriate performance values:

\begin{lstlisting}
def simulate(n, t):
    """
    Simulate the system of n philosophers for up to t time units.
    Return the average waiting time.
    """
    env = simpy.Environment()
    chopsticks = [simpy.Resource(env, capacity=1) for i in range(n)]
    philosophers = [
        Philosopher(env, chopsticks[i], chopsticks[(i + 1) % n], i) 
        for i in range(n)]
    env.run(until=t)
    return sum(ph.waiting for ph in philosophers) / n
\end{lstlisting}

This all-in-one function will be called for different values of \ic{n},
and the results will be plotted with \ic{matplotlib}---the ubiquitous Python
plotting engine.

\begin{lstlisting}
# Set up the plotting system
import matplotlib.pyplot as plt
import matplotlib
matplotlib.style.use("ggplot")

# Simulate
N = 20
X = range(2, N)
Y = [simulate(n, 50000) for n in X]

# Plot
plt.plot(X, Y, "-o")
plt.ylabel("Waiting time")
plt.xlabel("Number of philosophers")
plt.show()
\end{lstlisting}

\begin{figure}[tb]\centering
  \includegraphics[width=\columnwidth]{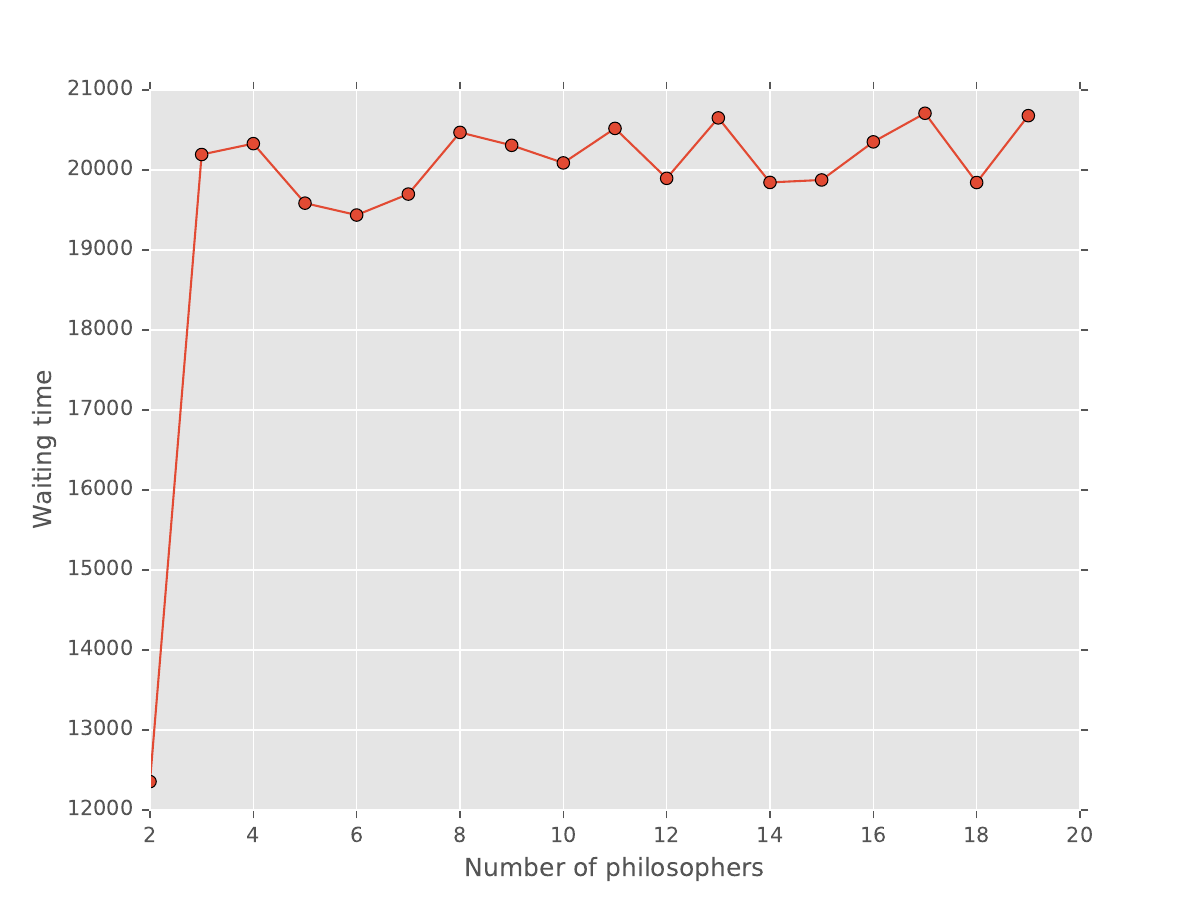}
  \caption{\label{waiting}Waiting time vs. the number of philosophers}
  \Description{Waiting time vs. the number of philosophers}
\end{figure}

Figure~\ref{waiting} shows the simulation results. As expected, the
waiting time for two philosophers is shorter than for three. A less
obvious result is that the waiting time for three or more philosophers
is almost constant with respect to the party size.

Any interpretation of these facts is beyond the scope of this story.

\section{Adding a Container}

No matter how you feel about the stubborn philosophers, the bottomless
bowl of rice at the center of the table is unrealistic. Real-life rice
ends soon because it is a {\em consumable resource}. The model can be
made less fictitious by treating the rice bowl as a container with
finite capacity. The amount of rice in the bowl will decrease by a
fixed amount every time a philosopher eats from it. (For convenience,
let's assume that the rice is atomically transferred to the
philosopher's plate when he collects both chopsticks.) Let's call the
new breed of philosophers ``pseudo-philosophers'' because genuine
philosophers surely do not eat rice or anything material but merely
pretend to be eating.

A \ic{simpy.Container} can yield two events of interest:
\ic{put(amount)} and \ic{get(amount)}. Both can block the generator if
the container is nearly full or nearly empty, respectively. A
container's capacity and current level are stored in the namesake
attributes, \ic{container.capacity} and \ic{container.level}.

Since a \ic{Philosopher} class already exists, the cheapest way to
implement a pseudo-philosopher is to add an optional \ic{bowl}
parameter to the constructor and a request for a portion of rice to
the \ic{get\_hunry()} method if the bowl has been provided (the
highlighted lines are responsible for the interaction with the
container):

\begin{lstlisting}
class Philosopher():
    T0 = 10 # Mean thinking time
    T1 = 10 # Mean eating time
    DT = 1 # Time to pick the other chopstick
    # START_HIGHLIGHT
    PORTION = 20 # Single meal size
    # END_HIGHLIGHT

    # START_HIGHLIGHT
    def __init__(self, env, chopsticks, my_id, bowl = None, DIAG = False):
    # END_HIGHLIGHT
        self.env = env
        self.chopsticks = sorted(chopsticks, key=id)
        self.id = my_id
        self.waiting = 0
        # START_HIGHLIGHT
        self.bowl = bowl
        # END_HIGHLIGHT
        self.DIAG = DIAG
	# Register the process with the environment
        env.process(self.run_the_party())

    def get_hungry(self):
        start_waiting = self.env.now
        self.diag("requested chopstick")
        rq1 = self.chopsticks[0].request()
        yield rq1 
        
        self.diag("obtained chopstick")
        yield self.env.timeout(self.DT)

        self.diag("requested another chopstick")
        rq2 = self.chopsticks[1].request()
        yield rq2

        self.diag("obtained another chopstick")
        # START_HIGHLIGHT
        if self.bowl is not None:
            yield self.bowl.get(self.PORTION)
            self.diag("reserved food")
        # END_HIGHLIGHT

        self.waiting += self.env.now - start_waiting
        return rq1, rq2

    def run_the_party(self):
        ...
\end{lstlisting}

Incidentally, the new class design allows me to feed different
philosophers from different bowls and even mix ``true philosophers''
and ``pseudo-philosophers.''

If the philosophers continue eating their rice, the bowl eventually
becomes empty, and the party ends. Another process, a chef, is solely
responsible for replenishing the bowl every \ic{T2} time units:

\begin{lstlisting}
class Chef():
    T2 = 150
    def __init__(self, env, bowl):
        self.env = env
        self.bowl = bowl
        env.process(self.replenish())

    def replenish(self):
       while True:
           yield self.env.timeout(self.T2)
           if self.bowl.level < self.bowl.capacity:
               yield self.bowl.put(self.bowl.capacity - self.bowl.level)
\end{lstlisting}

The updated function \ic{simulate()} has additional code for
instantiating the bowl and the chef and passes the \ic{bowl} parameter
to the philosophers:

\begin{lstlisting}
def simulate(n, t):
    """
    Simulate the system of n philosophers for up to t time units.
    Return the average waiting time.
    """
    env = simpy.Environment()
    # START_HIGHLIGHT
    rice_bowl = simpy.Container(env, init=1000, capacity=1000)
    chef = Chef(env, rice_bowl)
    # END_HIGHLIGHT
    chopsticks = [simpy.Resource(env, capacity=1) for i in range(n)]
    philosophers = [
        Philosopher(env, (chopsticks[i], chopsticks[(i + 1) % n]), i, 
        # START_HIGHLIGHT
                    rice_bowl) 
        # END_HIGHLIGHT
        for i in range(n)]
    env.run(until=t)
    return sum(ph.waiting for ph in philosophers) / n
\end{lstlisting}

\begin{figure}[tb]\centering
  \includegraphics[width=\columnwidth]{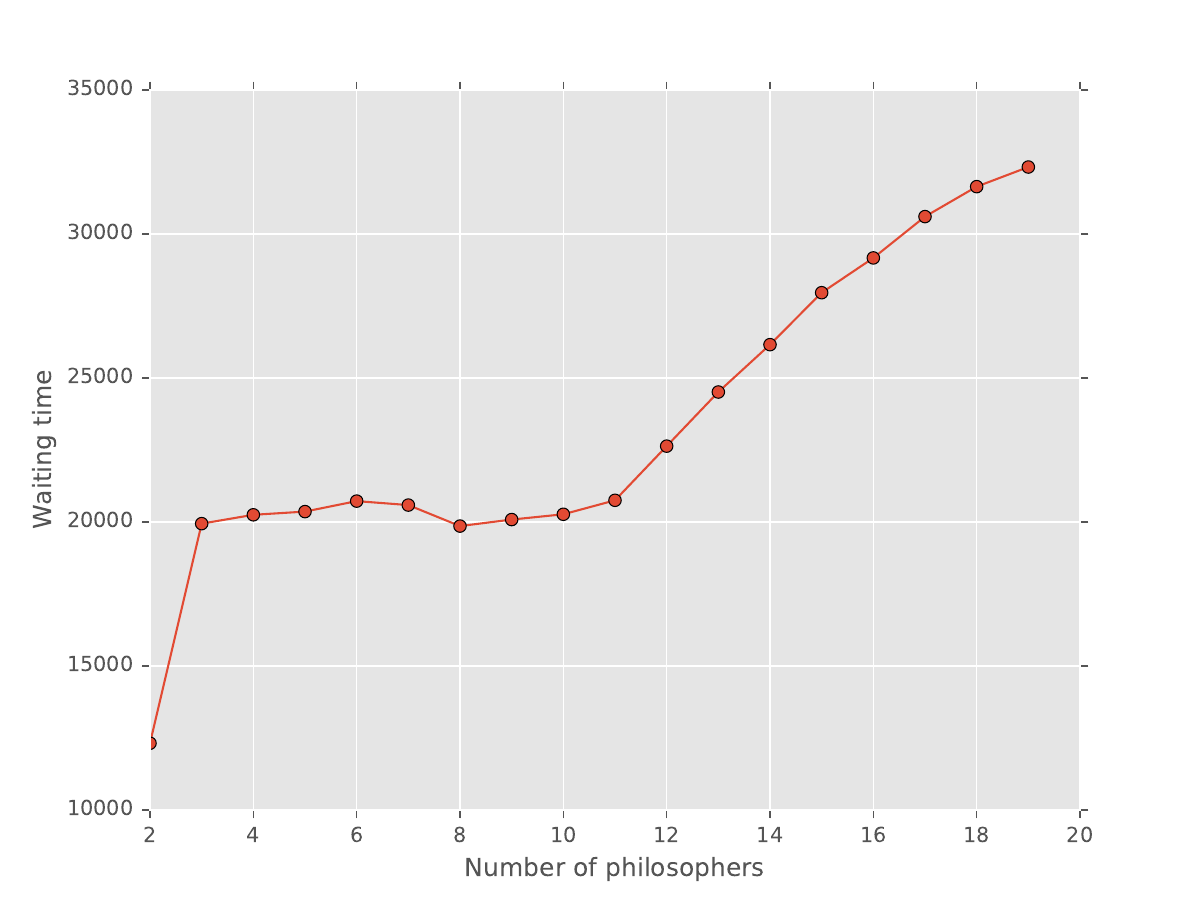}
  \caption{\label{wait_vs_n}Waiting time vs. the number of
    pseudo-philosophers} \Description{Waiting time vs. the number of
    pseudo-philosophers}
\end{figure}

Figure~\ref{wait_vs_n} shows the waiting time for the new realistic
model. As expected, there is almost no difference between the two
models when the number of dining philosophers is small (less than ten
in this case): the chef is cooking fast enough to supply the food. When
the number of partygoers increases, the turnaround time of the chef
becomes an issue and negatively affects the waiting time of the
patrons.

\section{Learning to Give Up}

A hungry philosopher may not be that hungry, after all. If the central
bowl has no rice and the chef is nowhere to be found, the philosopher
may call it a day: put down the chopsticks and return to
thinking. Presumably, he will have a chance to eat a supersized meal
next time.

Resource requests are generally blocking. They can block the caller
forever if the resource never becomes available (this is the reason
for deadlocks). \ic{SimPy} provides a mechanism for restricting the
wait time by failing events and waiting for multiple events.

Introducing the new---impatient---kind of philosophers requires
changing the methods \ic{get\_hungry()} and
\ic{run\_the\_party()}. First, a constant \ic{MAX\_WAIT} is added that
limits the waiting time to at most half of the time required to
replenish the rice bowl. Second, when a philosopher collects both
chopsticks, the model fires two events: the container request (as
before) and a timeout. The generator process is unblocked when {\em
  any} of these events trigger, which is accomplished by combining the
events with the logical ``or'' operator \ic{|} (a vertical bar). The
operator is a shortcut to the function
\ic{simpy.events.AnyOf(env,events)}. If the modeling logic requires
triggering {\em all} events before unblocking the process, use the
logical ``and'' operator \ic{\&} or the function
\ic{simpy.events.AllOf(env,events)} instead\footnote{If you are
familiar with UML, note that \ic{|} corresponds to the UML \ic{merge}
and \ic{\&} corresponds to the UML \ic{join}.}.

\begin{lstlisting}
class Philosopher():
    ...
        # START_HIGHLIGHT
    MAX_WAIT = Chef.T2 / 2
        # END_HIGHLIGHT
    ...

    def get_hungry(self, meal_size):
        start_waiting = self.env.now
        self.diag("requested chopstick")
        rq1 = self.chopsticks[0].request()
        yield rq1 
        
        self.diag("obtained chopstick")
        yield self.env.timeout(self.DT)

        self.diag("requested another chopstick")
        rq2 = self.chopsticks[1].request()
        yield rq2

        self.diag("obtained another chopstick")
        if self.bowl is not None:
            request = self.bowl.get(meal_size)
        # START_HIGHLIGHT
            yield request | self.env.timeout(self.MAX_WAIT)
            if request.processed:
                self.diag("reserved food")
            else: # Timeout
                self.diag("gave up")
                self.waiting += self.env.now - start_waiting
                yield simpy.Event(self.env).fail(ValueError(rq1, rq2))
        # END_HIGHLIGHT
                
        # Either no bowl or no timeout!
        self.waiting += self.env.now - start_waiting
        return rq1, rq2
\end{lstlisting}

After yielding the compound event, the program if \ic{SimPy} processed
the container request by looking at the attribute
\ic{request.processed}. If it did, the food was reserved, and the
philosopher could start eating. Otherwise, the wait has been
interrupted by the timeout, and \ic{get\_hungry()} must inform the
parent process of the failure. \ic{SimPy} allows a process to fail
(and raise an exception) by yielding a namesake event, as on the last
highlighted line of the previous listing. Incidentally, the raised
\ic{ValueError} exception takes the chopstick request handles as the
parameters: when \ic{get\_hungry()} fails, there is no other way to
return the handles.

The \ic{run\_the\_party()} method is instrumented with simple meal
size accounting. Whenever a philosopher chooses not to wait for the
chef, his next rice allowance is increased by one standard portion
size. (Intuitively, his overall wait time increases because some
attempts are futile). Failed attempts are detected; the failures are
mitigated by catching the exception raised by \ic{get\_hungry()} and
extracting the chopstick request handles from the exception.

\begin{lstlisting}
def run_the_party(self):
    # START_HIGHLIGHT
    meal_size = self.PORTION
    # END_HIGHLIGHT
    while True:
        yield self.env.timeout(random.expovariate(1 / self.T0))

        get_hungry_p = self.env.process(self.get_hungry(meal_size))
    # START_HIGHLIGHT
        try:
            rq1, rq2 = yield get_hungry_p 
            yield self.env.timeout(random.expovariate(1 / self.T1))
            meal_size = self.PORTION
        except ValueError as values: # Timeout
            rq1, rq2 = values.args
            meal_size += self.PORTION
    # END_HIGHLIGHT

        self.chopsticks[0].release(rq1)
        self.chopsticks[1].release(rq2)
        self.diag("released chopsticks")
\end{lstlisting}

The new timeout feature can be disabled by setting \ic{MAX\_WAIT} to a
substantially large number---say, larger than \ic{T2}.

\begin{figure}[tb]\centering
  \includegraphics[width=\columnwidth]{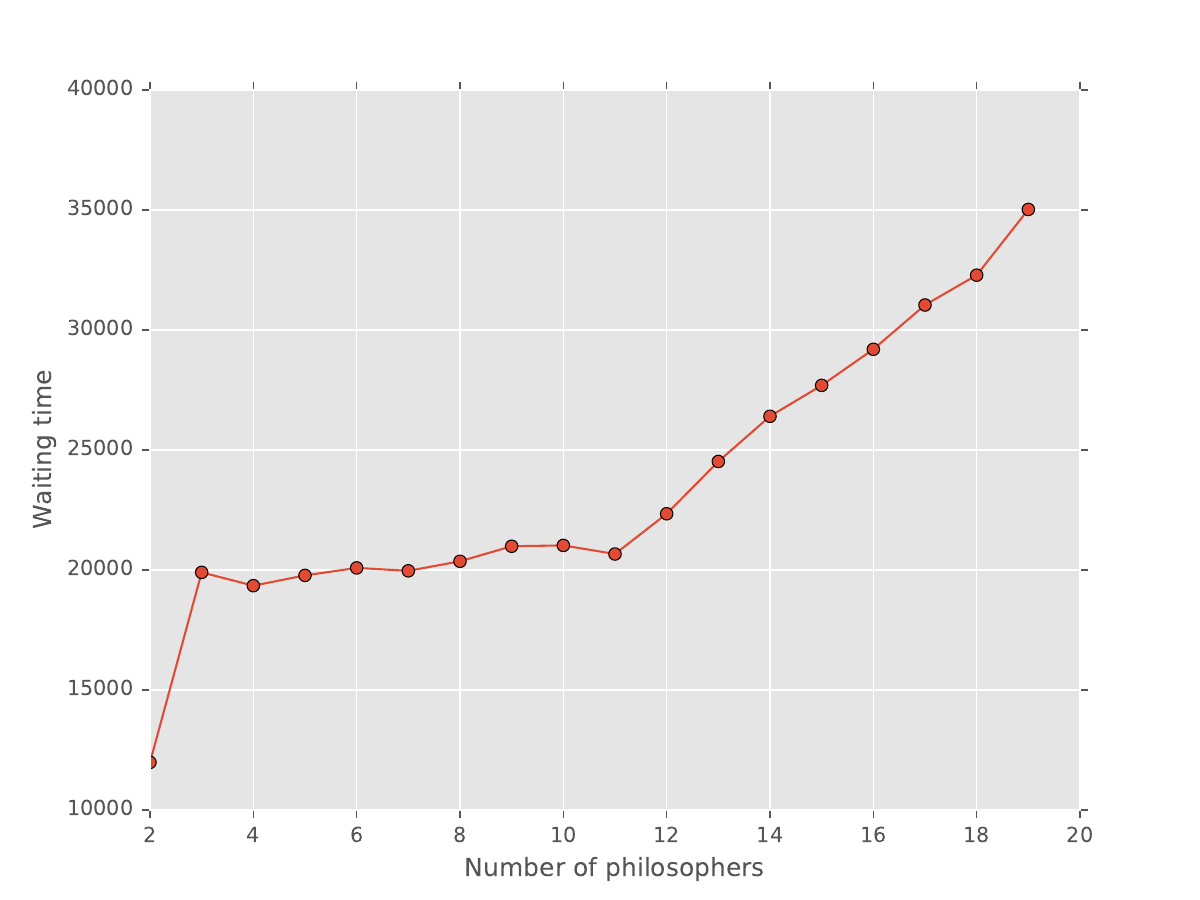}
  \caption{\label{wait_vs_imp}Waiting time vs. the number of impatient
    philosophers} \Description{Waiting time vs. the number of
    impatient philosophers}
\end{figure}

Figure~\ref{wait_vs_imp} shows the waiting time for the impatient
philosophers. The difference between this model and the previous one
is visible only when the number of partygoers is large enough.

\section{Learning to Communicate}

Another classical example---a customer service counter---explains how
to model processes that interact directly rather than by sharing
resources. In the new scenario, customers arrive with exponentially
distributed delays and queueing into a line. (In the example below,
the first ten customers are conveniently ``generated'' by a
\ic{customer\_generator\_p()} process.)  The operator at the customer
service counter takes the next customer from the line and services
them in exactly \ic{SERVICE\_DELAY} time units. A customer is served
successfully with a probability of 0.9. If there are no customers in
the line, the operator becomes idle (``falls asleep,'' if you like)
but is reactivated (``woken up'') by the next arriving customer.

\ic{SimPy} provides two direct interaction mechanisms: process
interruption and event-based synchronization. To interact, \ic{SimPy}
processes must have references to either each
\ic{simpy.events.Process}, as returned by calls to \ic{env.process()},
or to the events involved in the interaction
(\ic{simpy.events.Event}).

The following code illustrates both communication mechanisms. It
begins with the imports and definitions of global variables. The
Python standard double-ended queue \ic{collections.deque} models the
service line.

\ic{SimPy} uses Python exceptions to manage process interruptions. A
new exception class, \ic{CustomerFa\-iledException}, is created to
distinguish \ic{SimPy} process-related exceptions from proper Python
exceptions.

\begin{lstlisting}
import random
from collections import deque
import simpy

# Shared global variables and "constants"
SERVICE_DELAY = 10
service_line = deque()
counter_idle = False # The state of the operator at the service counter
env = simpy.Environment()

class CustomerFailedException(Exception):
    pass
\end{lstlisting}

The model consists of three types of processes. The
\ic{customer\_generator()} function imitates the rest of the world. In
a complete model, customers would arrive from other model components
rather than being ``born'' on the spot.

\begin{lstlisting}
def customer_generator():
    for _ in range(10):
        env.process(customer())
        yield env.timeout(random.expovariate(1 / SERVICE_DELAY))

customer_generator_p = env.process(customer_generator())
\end{lstlisting}

Customers in the model are simple stateless creatures. Once born, they
obtain a service ticket represented as a \ic{simpy.events.Event} and
enter it into the \ic{service\_line}. If the service counter operator is
asleep (\ic{counter\_idle==True}), the customer interrupts the counter
process by calling \ic{counter\_p.interrupt()}. This method
raises a \ic{simpy.Interrupt} exception in the target process and
``awakens'' it. The matching lines are highlighted in the code below.

Now that the operator is not asleep anymore, the customer \ic{yield}s the
ticket. The event gets {\em triggered}. The customer is blocked until
some other process processes the event. Naturally, the event becomes
{\em processed}, and if it is processed successfully, the execution of
the customer resumes\footnote{If you are familiar with the C
\ic{pthreads}, you may recognize a striking similarity between
\ic{SimPy} events and \ic{pthread\_cond\_t}. Triggering and processing
an event are loose equivalents of \ic{pthread\_cond\_wait()} and
\ic{pthread\_cond\_signal()}.}. Otherwise, a
\ic{CustomerFailedException} is raised, which terminates the simulator
if it is not correctly handled.

\begin{lstlisting}
def customer():
    print("Customer arrived @{0:.1f}".format(env.now))
    ticket = env.event()
    service_line.append(ticket)

    if counter_idle:
	# START_HIGHLIGHT
        counter_p.interrupt()
	# END_HIGHLIGHT

    try:
        yield ticket
        print("Customer left @{0:.1f}".format(env.now))
    except CustomerFailedException:
        print("Customer failed (and left) @{0:.1f}".format(env.now))
\end{lstlisting}

The service counter is a potentially infinite process. If there is at
least one waiting customer, the first ticket is removed from the
service line, and the corresponding customer is taken care of. Then
the counter process ``flips a coin'' and either fails the ticket event
(\ic{ticket.fail("CustomerFailedException")}) or succeeds it
(\ic{ticket.succeed()}). In the former case, a
\ic{CustomerFailedException} is raised on the client side, as
explained above.

If the service line is empty, the counter operator becomes idle and
\ic{yield}s an event. This event is triggered but never processed,
putting the operator in eternal sleep. However, as explained above, sleep
can also be interrupted by an arriving customer process.

\begin{lstlisting}
def counter():
    global counter_idle
    while True:
        if service_line:
            ticket = service_line.popleft()
            yield env.timeout(SERVICE_DELAY)
            if random.randint(0,9) == 9:
                ticket.fail(CustomerFailedException())
            else:
                ticket.succeed()
        else:
            counter_idle = True
            print("The operator fell asleep @{0:.1f}".format(env.now))
            try:
                yield env.event()
	# START_HIGHLIGHT
            except simpy.Interrupt:
	# END_HIGHLIGHT
                counter_idle = False
                print("The operator woke up @{0:.1f}".format(env.now))

counter_p = env.process(counter())
\end{lstlisting}

The simulation's sample output follows:

\begin{lstlisting}
env.run()
#> The operator fell asleep @0.0
#> Customer arrived @0.0
#> The operator woke up @0.0
#> Customer arrived @0.4
#> Customer left @10.0
#> The operator fell asleep @20.0
#> Customer left @20.0
#> Customer arrived @24.6
#> The operator woke up @24.6
#> The operator fell asleep @34.6
#> Customer left @34.6
#> Customer arrived @59.3
#> The operator woke up @59.3
#> Customer arrived @64.2
#> Customer arrived @65.9
#> Customer left @69.3
#> Customer arrived @75.8
#> Customer left @79.3
#> Customer arrived @83.1
#> Customer arrived @86.7
#> Customer left @89.3
#> Customer arrived @90.9
#> Customer left @99.3
#> Customer left @109.3
#> Customer left @119.3
#> The operator fell asleep @129.3
#> Customer left @129.3
\end{lstlisting}

You can further instrument the model with statistics-gathering tools.

\section{Conclusion}

This article demonstrates the possibility of using Python for
not-so-trivial M\&S projects. The ``M'' (modeling) part of the story
is somewhat sketchy. Hovewer, computer modeling, even discrete event
modeling alone, is too vast to be covered in a dozen pages without
distracting the reader's attention from the holistic M\&S experience.

On the \ic{SimPy} side, the most notable features that should be
sufficient for many simple M\&S projects were reviewed. Here is a
summary of what was left uncovered:

\begin{itemize}
\item \ic{env.schedule(event, priority=1, delay=0)}---A mechanism for
  scheduling an event in the future rather than triggering it
  immediately.
\item \ic{event.callbacks}---A list of callback
  functions that will be called when the event is processed. (The
  default callback always resumes the blocked process that triggered
  the event.)
\item \ic{PriorityResource},
  \ic{PreemptiveResource}, \ic{PriorityStore},
  \ic{FilterStore}---A collection of specialized resources.
\item \ic{RealtimeEnvironment}---An environment for
  real-time simulation that requires synchronization between the
  simulation clock and the ``wall clock.''
\end{itemize}

You are referred to the \ic{SimPy} documentation~\cite{simpy} for
additional information.

\bibliographystyle{acm}
\bibliography{simpy}

\end{document}